# ENHANCED SPREADSHEET COMPUTING WITH FINITE-DOMAIN CONSTRAINT SATISFACTION

Ezana N. Beyenne and Hai-Feng Guo

Department of Computer Science, University of Nebraska at Omaha
Omaha, NE 68182-0500, USA

## ABSTRACT

*The spreadsheet application is among the most widely used computing tools in the modern society. It provides great usability and usefulness, and it easily enables a non-programmer to perform programming-like tasks in a visual tabular "pen and paper" approach. However, due to its mono-directional dataflow, spreadsheets are mostly limited to bookkeeping-like applications. This paper shows how the spreadsheet computing paradigm is extended to break through this limitation for solving constraint satisfaction problems. We present an enhanced spreadsheet system where finite-domain constraint solving is well supported in a visual environment. A spreadsheet-specific constraint language is constructed for general users to specify constraints among data cells in a declarative and scalable way. The new spreadsheet system significantly simplifies the development of many constraint-based applications using a visual tabular interface. Examples are given to illustrate the usability and usefulness of the extended spreadsheet paradigm.*

## 1.INTRODUCTION

Due to its great usability and usefulness, the spreadsheet application has been widely used in the modern computing society. It easily enables a nonprogrammer perform programming-like tasks, such as personnel information management, computerized bookkeeping, and electronic financial planning, in a tabular "pen and paper" approach. Each cell in a spreadsheet may contain a value in some data type or a formula which depends on other cells; then when the other cells are updated, the dependent cell is automatically updated as well. Such a mono-directional dataflow makes the spreadsheets fairly easy for general users to understand and manipulate. However, it also hampers the further uses of spreadsheets on many daily-life constraint problems, such as resource allocation, task scheduling, and timetabling problems.

A few research efforts have been conducted to extend spreadsheets with more powerful computational tools for solving domain-specific problems. For instance, the spreadsheet interface for sales/products configuration system [7] was developed by incorporating a light-weight constraint satisfaction solver (ILOG Solver) as an MS Excel add-on. PrediCalc [12] is a spreadsheet system which allows general logical constraints for data management systems; logical constraints, constructed using cell names and the usual logical connectives and the quantifiers, are used to maintain the consistency among databases. IntervalSolver [10] is a spreadsheet extension that allows users to take uncertainties of values into account by using intervals; it is capable of dealing with rounding errors, imprecise data and numerical constraints on intervals.

      

More general extensions combine a spreadsheet paradigm with other programming paradigms, such as constraint logic programming. PERPLEX [2] is a programming environment combining the power of logic programming with the spreadsheet interface; it allows users to define predicates among data cells in an interactive way. NExSched [6] is a spreadsheet interface for solving constraint satisfaction problems following the knowledge-sheet paradigm [9]; it incorporates a finite-domain constraint solver (SICStus) as a plug-in for Microsoft Excel. NEXCEL [4, 5, 15] proposed a deductive extension of the spreadsheet paradigm for reasoning and decision-making applications; it allows users to define logical inference rules for symbolic reasoning, and values satisfying them are visualized on the fly and propagated to any cells which depend on them.

All of these extensions essentially provide a spreadsheet interface for constraint (logic) programming or solving domain-specific problems, which require that the defined predicates, constraints, and inference rules follow the correct syntax of underlying programming paradigm. In other words, it requires users to have certain programming skills or experiences on solving domain-specific problems. Those extensions certainly improve the usefulness of spreadsheets; however, without improving the usability at the same time, all extensions are nearly inaccessible to mainstream spreadsheet users. From this point of view, unless the extensions attract mainstream spreadsheet users, the practical contribution of these extensions is limited since domain experts would rather use an underlying programming paradigm directly. For this reason, it is very important to have a user-centric approach to functions in Excel. As emphasized in [11] by Jones, et al., "the commercial success of spreadsheets is largely due to the fact that many people find them more usable than programming languages for programming-like tasks."

In this paper, we present a development framework that incorporates spreadsheets with a finite-domain constraint programming paradigm, where each spreadsheet cell can be attached to a finite domain or a constraint specifying the relation among cells. For an experienced programmer, scalability issues are usually handled nicely by either recursion or loop statements. However, for a general spreadsheet user, a larger-size problem usually requires large number of constraints to be specified without the assistance of recursion or loops. This can make the spreadsheet interface impractical and quite error-prone for constraint programming. Therefore, we provide a set of new functions for spreadsheet users to specify the constraints among data cells in a declarative and scalable way. The new framework significantly simplifies the development of many constraint-based applications using a visual spreadsheet interface.

We developed an extension tool, named CSeE (Constraint Satisfaction extended Excel), between Microsoft Excel and SWI Prolog. This work focuses on merging the CLP(FD)[1] paradigm with the Excel's spreadsheet paradigm in such a way that it hides the complexities of CLP(FD) by providing an add-in to Excel in the form of menus and function extensions. These menus and functions will enable the end users to specify constraints that will interact with a back-end constraint solver (SWI Prolog). The general user will be able to use the extended spreadsheet to explore new ways of solving many problems (e.g., scheduling problems, puzzle solving, etc.) that have previously proved to be an arduous, time-consuming task and also were inaccessible due to syntactic and semantic complexities. The extended spreadsheet will become an even more indispensable tool in solving previously complex and error-prone problems that they encounter in their daily lives. The CSeE tool also provides an interface for an experienced constraint programmer to define new predicates and propagation rules. .
Recently, a new Excel solver [16, 17] has been proposed as an education platform for management science. The solver delivers great usability and quality on solving optimization, simulation, and decision science problems. Different from the Excel solver, our CSeE system applies an AI-based reasoning approach and multi-directional dataflow models [1, 13] to solve general finite-domain constraint satisfaction problems. Our system further provides a set of spreadsheet-specific constraints in the form of Excel functions to handle large problems in a declarative and scalable way.

The remaining parts of this paper are organized in the following way: Section 2 gives a brief introduction of constraint programming. Section 3 shows the architectural design of the CSeE system, which enhances the Microsoft Excel system with a spreadsheet-specific constraint language (SSCL) and constraint satisfaction capability. Section 4 introduces the syntax and informal semantics of the SSCL constraints. Section 5 explains how the constraints defined in the spreadsheet layer are integrated to an executable constraint logic program. Three constraint-based spreadsheet applications are given in Section 6 to illustrate the usability and usefulness of the CSeE tool. Finally, the conclusion and future extensions are addressed in Section 7.

## 2. CONSTRAINT PROGRAMMING

Constraint programming (CP) [1, 13] is a computational programming paradigm for formulating and solving problems based on constraints. A constraint is a description of the logical relation among several variables, where each variable takes values from a given domain. A constraint is therefore used to determine the values that variables can take, representing some partial information about related variables.

Constraints are a natural medium to express all sorts of daily reasoning problems. For example, the following constraints can be used to describe a relation among the three angles ($A_1$, $A_2$ and the right angle $A_3$) in a right triangle:

$$A_1 + A_2 + A_3 = 180 \text{ and } A_3 = 90,$$

and a simple loan can be represented as

$$B = P + I \times P,$$

where $B$, $I$, and $P$ are the final balance, the interest rate, and the principal of the loan, respectively. CP is an emergent software technology for specifying and solving many types of combinatorial and optimization problems in a declarative yet effective way. The idea of constraint programming is to model and solve problems by specifying constraints which must be satisfied in the problems. An important characteristic of constraint programming is its declarative nature. A CP programmer does not need to specify the computational procedure that should be used to enforce the constraints; rather, they just state what constraints must hold and leave them up to a computational system to solve.

In this paper, we focus on an important domain of constraints, called finite-domain constraints, since probably more than 95% of the current industrial constraint applications use finite domains [1]. Constraint Satisfaction is an algorithmic study of solving finite-domain constraints by the Artificial Intelligence community. A Constraint Satisfaction Problem (CSP) can be formally defined as

- **Variables**: a set of variables $X = \{ x_1, \cdots, x_n \}$,
- **Domains**: for each variable $x_i$, a finite set $D_i$ of possible values, and

- **Constraints**: a set of constraints restricting the values that the variables can take simultaneously.

Constraint satisfaction is to find solutions of a CSP by assigning each variable a value from its domain, which satisfy all the constraints. Often there are multiple solutions that satisfy the constraints. In an optimization problem, one can specify a criterion to choose the optimal solution among the solution choices.

**Example 1. The Knapsack Problem:** A smuggler has a knapsack that has limited capacity of only 9 units. He can smuggle bottles of whiskey of consisting of 4 units, perfume bottles of 3 units and 2 unit cartons of cigarettes. The profit from smuggling a bottle of whiskey is $15, while a bottle of perfume is $10 or a carton of cigarettes is $7. The smuggler can only make the trip if he can make a profit of greater than or equal to $30. What are the combinations that will maximize his profit?

A constraint program, solving the Knapsack problem, is shown as follows:

domain([W,P,C], 0, 9),                                     (1)
4*W + 3*P + 2*C <= 9,                                       (2)
15*W + 10*P + 7*C >= 30,                                    (3)
maximize(labeling([W,P,C]), 15*W + 10*P + 7*C).             (4)

The structure of the program follows the definition of a constraint satisfaction problem. Line (1) introduces a set of variables *{W, P, C}*, denoting the quantities of whiskey bottles, perfume bottles, and cigarette cartons to smuggle and also defines their domains, an integral range from 0 to 9. The rest of code specifies the constraints declared in the problem, where the last line is an optimization predicate to search the value space of *{W, P, C}* to maximize the profit objective, $15 * W + 10 * P + 7 * C$. The built-in predicate labeling(L) assigns concrete possible values to the variables in *L* from their given domains.

To find more detailed introduction and algorithms on constraint satisfaction, please refer to [1, 13].

## 3. THE CSeE SYSTEM — DESIGN ARCHITECTURE

The CSeE system enhances the Microsoft Excel with a spreadsheet-specific constraint language (SSCL) and constraint satisfaction capability. The enhancement is currently implemented as a COM Add-In for MS Excel 2003 developed using the C# and Visual Studio Tools for Office (VSTO) 2005. The spreadsheet-specific constraint language (SSCL) is a set of easy-to-use Excel functions, which can easily be scaled to specify large problems in a consistent form. We use the SWI-Prolog as the back-end constraint solver. The bi-directional communication between the Excel and SWI-Prolog is achieved by language processing and transformation techniques.

The overall architecture of our CSeE system is shown in Figure 1. The user-centric design of the architecture focuses on both the usability and usefulness of the system. The new Excel spreadsheet provides menus, submenus, and access to spreadsheet-specific constraint language (SSCL), in a form of Excel functions, to allow users to model and program constraint satisfaction problems in a scalable yet similar visual manner of using Microsoft Excel. The language processing layer parses the spreadsheet cell inputs, compiles to the objective CLP code, and integrates the compiled code with a pre-defined CSeE library for SWI-Prolog to generate an executable CLP program. Additionally, the language processing layer decomposes the query result, which is returned from the constraint solver, into cell values, so that the spreadsheet will be properly updated with solutions.

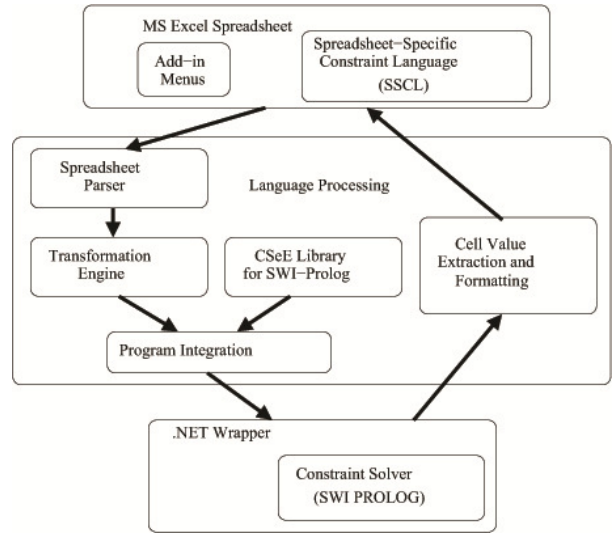

Figure 1: The CSeE System Architecture

The constraint solver layer is comprised of a .NET wrapper, named CSeEWrapper, which acts as the communication channel between the SWI Prolog and other applications in the .NET framework. The CSeEWrapper is a class that wraps the SWI Prolog's plcon.exe window using the .NET 2.0's System.Diagnostics.Process namespace, thereby enabling the CSeE System to interface with SWI-Prolog. Rich class libraries, templates, languages and technologies make the .NET framework an ideal platform for developing both the communication interface and Office applications. In addition to reusability, the framework allows developers to use a multitude of languages, such as C# or VB.NET, to manipulate the Office application's object model and interface with third party software in a seamless managed environment. One such library provided by the .NET Framework's is the System Diagnostics library, which is used to wrap SWI-Prolog's command window. The library provides functionality to send data (e.g., programs generated from the parser, queries) and receive data (e.g., results from the parser) from SWI-Prolog's command window, in a synchronous mode. It also allows the SWI-Prolog functionality to be ported to other .NET applications.

## 4. SPREADSHEET-SPECIFIC CONSTRAINT LANGUAGE (SSCL)

Spreadsheets are a popular paradigm due to their ease of use and usefulness that has enabled non-programmers to perform programming like tasks in a familiar tabular "pen and paper" setting. For the same reason, we present a spreadsheet-specific constraint language (SSCL) for general users to specify constraint satisfaction problems over the Excel spreadsheets. The SSCL language provides familiar and concrete representations of domain variables and their values, as well as constraints on these domains in the form of functions. Typically, constraint logic programming relies on control statements, such as recursion, to handle large-scale problems. Without the aid of recursion, spreadsheet users will find that as the scale of the problem increases, a larger number of constraints must be specified to solve constraint satisfaction problems. The SSCL adds the benefit of hiding the syntactic and semantic complexity of constraint logic programming, while at the same time providing flexibility to specify large-scale applications. The SSCL introduces a set of constraint functions that will allow users to specify constraints that exist among cells in a declarative and scalable way.

The design of SSCL follows the basic principle of defining a constraint satisfaction problem, which includes a set of finite-domain variables and a finite set of constraints defined over them. Besides standard use of the spreadsheet cell, each cell can further be attached to

- a finite domain; a spreadsheet cell, which is attached to a finite domain, is often called a cell variable;
- a constraint describing the relationship that exists among other cell variables;
- an optimization constraint; or
- a variable/constraint range specification.

We describe the syntax and informal meaning of the SSCL constraints in the follow subsections. Each SSCL constraint has a prefix of *ss*, and its syntactic conventions in regard to the MS Excel cell range is consistent to Excel's A1 reference style [3], which refers to column headings starting with letters beginning from A through IV ( a total of 256 columns) and refers to row headings consisting of numbers from 1 through 55,536. For example, A1 refers to the cell in column A and row 1; and B5:E10 refers to the range of cells in columns B through E and rows 5 through 10.

## 4.1. Domain Specification

The finite domain can be specified individually on a cell (e.g., A1) as follows:

- an integer list (or an integer): The cell variable A1 can take a value in an integer list within a pair of square brackets (e.g., a list [2, 5, 7]).
- *Lower..Upper*: The cell variable A1 can take an integer value between *Lower* and *Upper*; for example, 2..5 means an integer list [2, 3, 4, 5].

The finite domain can also be specified for a cell range using a domain constraint, in a form of ssDomain(CellRange, Min, Max) or ssDomain(CellRange, IntegerList), where each cell variable in CellRange can take a value from the specified finite domain. For example, ssDomain(A1:H8, 0, 1) defines Boolean variables among cells A1 through H8.

## 4.2. Constraint Specification

A constraint describes a relation among cell variables that a solution must satisfy. Let $A1$ and $B2$ be two cell variables whose domains are set to 1..10. Thus, $A1+4 < B2$ is a finite-domain arithmetic constraint. Using consistency techniques [1, 13] in the constraint solver SWI Prolog, the CSeE system is able to automatically reduce the domains of $A1$ and $B2$ to 1..5 and 6..10, respectively. The consistency techniques can discover and remove all of the invalid values. For example, 6 is invalid in $A1$'s domain because there is no value from the domain of $B2$ to support the constraint $6 + 4 < B2$. Therefore, 6 is removed from $A1$'s domain automatically.

### 4.2.1. Spreadsheet-Specific Constraints

Besides the regular relational constraints and symbolic arithmetic expressions supported directly by SWI-Prolog [14], the SSCL language provides a set of spreadsheet-specific constraints defined as follows:

- **ssNthElement(N, L, V)**: V is constrained to be equal to the N-th element of the list L, where V is a cell variable, and L is a list of integers or a cell variable range.
- **ssAllDifferent($C_{TL}$:$C_{BR}$)**: All cell variables in the cell range from the top left cell, $C_{TL}$, to the bottom right cell, $C_{BR}$, are pairwise non-equal.
- **ssRowsAllDifferent($C_{TL}$:$C_{BR}$)**: The cell variables in each same row, within the cell range $C_{TL}$:$C_{BR}$, are pairwise non-equal. For example, ssRowsAllDifferent(A1:C2) is equivalent to
-

$$A1 \neq B1, A1 \neq C1, B1 \neq C1,$$
$$A2 \neq B2, A2 \neq C2, B2 \neq C2.$$

- **ssColsAllDifferent($C_{TL}$:$C_{BR}$)**: The cell variables in each same column, within the cell range $C_{TL}$:$C_{BR}$, are pairwise non-equal.
- **ssDiagonalsAllDifferent($C_{TL}$:$C_{BR}$)**: The cell variables in each diagonal cells, within the cell range $C_{TL}$:$C_{BR}$, are pairwise non-equal. For example, the following constraint ssDiagonalsAllDifferent(A1:D3) is equivalent to

$$A2 \neq B3,$$
$$A1 \neq B2, A1 \neq C3, B2 \neq C3,$$
$$B1 \neq C2, B1 \neq D3, C2 \neq D3,$$
$$C1 \neq D2.$$

- **ssBackDiagonalsAllDifferent($C_{TL}$:$C_{BR}$)**: The cell variables in each same back-diagonal cells, within the cell range $C_{TL}$:$C_{BR}$, are pairwise non-equal.
- **ssRowsAggregate(ArithOp, $C_{TL}$:$C_{BR}$, RelOp, IntList)**: The values of all cell variables in *i*-th same row, within the range of $C_{TL}$:$C_{BR}$, are aggregated together using the binary arithmetic operator ArithOp, and the result has the binary relation RelOp with the *i*-th integer in the list IntList. For example, ssRowsAggregate(+, A1:C2, <, D1:D2) means that

$$A1 + B1 + C1 < D1, A2 + B2 + C2 < D2.$$

If all of the integers in the IntList are same, they can be replaced with a single integer. For example, ssRowsAggregate(+, A1:C2, <=, 10) means that

$$A1 + B1 + C1 \leq 10, A2 + B2 + C2 \leq 10.$$

- **ssColsAggregate(ArithOp, $C_{TL}$:$C_{BR}$, RelOp, IntList)**: The values of all cell variables in *i*-th same column, within the range of $C_{TL}$:$C_{BR}$, are aggregated together using the binary arithmetic operator ArithOp, and the result has the binary relation RelOp with the *i*-th integer in the list IntList.
- **ssDiagonalAggregate(ArithOp, $C_{TL}$:$C_{BR}$, RelOp, IntList)**: Similar to the previous constraint, except that ssDiagonalAggregate/4 applies aggregation on each same diagonal cells. For example, ssDiagonalAggregate(+, $A1 : C2$, >, [2,5,4,6]) is equivalent to

$$A2 > 2, A1 + B2 > 5, B1 + C2 > 4, C1 > 6.$$

- **ssBackDiagonalAggregate(ArithOp, $C_{TL}$:$C_{BR}$, RelOp, IntList)**: Similar to the previous constraint, except that ssBackDiagonalAggregate/4 applies aggregation on each same back-diagonal cells.
- **ssPairsOp(C1$_{TL}$:C1$_{BR}$, ArithOp, C2$_{TL}$:C2$_{BR}$, RelOp, IntList)**: This constraint is used to apply the binary operator ArithOp on each pairwise *i*-th operands from the cell ranges C1$_{TL}$:C1$_{BR}$ and $C2_{TL}$:$C2_{BR}$, and the result has the binary relation RelOp with the *i*-th integer in the list IntList. For example, ssPairsOp(A1:B2, *, C1:D2, =, E1:F2) equals to
$$A1 * C1 = E1, B1 * D1 = F1, A2 * C2 = E2, B2 * D2 = F2.$$

#### 4.2.2. Optimization Constraints

- **ssMin(V)**: The minimal optimization constraint is used to find a solution to the constraint problem, where the value of the cell variable V is minimized.
- **ssMax(V)**: The maximal optimization constraint is used to find a solution to the constraint problem, where the value of the cell variable V is minimized.

#### 4.2.3. Variable/Constraint Range Specification

Users are required to specify the ranges of the cell variables and constraints, so that the CSeE system could easily identify all the finite-domain cell variables and their related constraints. The absence of these specifications will require the parsing of the entire spreadsheet or special indication of each constraint (e.g., within a pair of squared brackets); however, both alternatives could cause unnecessary ambiguous issues. Hence, we introduce two extra constraints as follows:

- **ssVarRange(CellRange)**: The cell variables are located in CellRange, where multiple ranges can be put together; e.g., ssVarRange([A1:C3,E6:F9,G10]).
- **ssConstraintRange(CellRange)**: The constraints are located in CellRange.

## 5. Language Processing Layer

The spreadsheet language processing layer gathers all the necessary input information (namely, cell variables, domains and constraints), parses through the Excel's A1 reference style [3], transforms the spreadsheet-specific constraints into SWI-Prolog's syntax, and integrates with the pre-defined CSeE library for SWI-Prolog into an executable CLP program. The language processing layer is also responsible for extracting and formatting values for each cell variable once the answers are returned from the constraints solver. In this section, we mainly address the transformation rules, the CSeE library, and program integration in details.

### 5.1. Cell Range Transformation

We first introduce transformation rules to convert a cell range in Excel's styles into a list of variables. As shown in Figure 2, given an input of a cell range *C*, the function varList(C) creates a list *L* of cell variables, and the function varListByRow(C), where C is in the form of $C_{TL}$ : $C_{BR}$, creates a list *L* containing sub-lists of cell variables grouped by rows. For example, varListByRow(A1:B3) returns [[A1,B1],[A2,B2],[A3,B3]]. Similarly, the following transformation functions are defined as well:

- **varListByCol($C_{TL}$ : $C_{BR}$)** creates a list containing sub-lists of cell variables grouped by columns.
- **varListByDiag($C_{TL}$ : $C_{BR}$)** creates a list containing sub-lists of cell variables grouped by diagonals. For example, the constraint varListByDiag(A1:B3) returns a nested list [[A3],[A2,B3],[A1,B2],[B1]].

International Journal of Programming Languages and Applications ( IJPLA ) Vol.5, No.2, April 2015International Journal of Programming Languages and Applications ( IJPLA ) Vol.5, No.2, April 2015

- **varListByBackDiag**($C_{TL} : C_{BR}$) creates a list containing sub-lists of cell variables grouped by back diagonals. For example, the constraint varListByBackDia(A1:B3) returns a list [[A1],[A2,B1],[A3,B2],[B3]].

```
varList(C)                              varListByRow(C_TL : C_BR)
if C is in form of C_TL : C_BR then     L ← [ ]
  L ← [ ]                               for r ← rowId(C_TL) to rowId(C_BR)
  for r ← rowId(C_TL) to rowId(C_BR)      L1 ← [ ]
    for c ← colId(C_TL) to colId(C_BR)    for c ← colId(C_TL) to colId(C_BR)
      L ← add(cellVar(c,r), L)              L1 ← add(cellVar(c,r), L1)
else if C is in form of [...] then      L ← add(L1, L)
  L ← C                                 OUTPUT: L
else
  L ← [C]
OUTPUT: L
```

Figure 2: Algorithms of varList and varListByRow

### 5.2. Constraint Transformation

In this subsection, we describe the major transformation rules, itemized as follows, from the spreadsheet cell inputs to the SWI-equivalent constraints. The transformation function is described in a form of

$$\frac{\text{Intermediate Steps}}{\text{Cell Input} \Rightarrow \text{SWI-equivalent Constraint}}$$

where Intermediate Steps are the auxiliary functions to make the transformation happen. We focus on those transformation rules which are particularly related to spreadsheets, and the rules for those constraint inputs which are not listed below are defined in a similar way.

- **domain2swi** converts a domain specification ssDomain to a SWI constraint using the following two rules, where ← means assignment. The finite domain (e.g., a single integer V, or a range Min..Max) can also be specified individually on a cell C. In such cases, constraints (e.g., C = V or C in Min..Max) for each cell C will be created.

$$\frac{\text{CL} \leftarrow \text{varList(C)}}{\text{ssDomain(C,IL)} \Rightarrow \text{CL in IL}} \qquad \frac{\text{CL} \leftarrow \text{varList(C)}}{\text{ssDomain(C,Min,Max)} \Rightarrow \text{CL in Min..Max}}$$

- **allRowDiff2swi** converts the spreadsheet constraint ssRowsAllDifferent to an SWI equivalent as follows, where the predicate subListAllDifferent/1, pre-defined in the CSeE library for SWI-Prolog, means that the values for the cell variables in each sublist are all different.

$$\frac{\text{L} \leftarrow \text{varListByRow}(C_{TL} : C_{BR})}{\text{ssRowsAllDifferent}(C_{TL} : C_{BR}) \Rightarrow \text{subListAllDifferent(L)}}$$

- **allDiaDiff2swi** converts the constraint ssDiagonalsAllDifferent in a similar way, except calling the rule varListByDiagonal. So are allColDiff2swi and allBDiaDiff2swi for converting constraints ssColsAllDifferent and ssBackDiagonalsAllDifferent, respectively, to the same SWI predicate subListAllDifferent/1.

$$\frac{L \leftarrow \text{varListByDiagonal}(C_{TL} : C_{BR})}{\text{ssDiagonalsAllDifferent}(C_{TL} : C_{BR}) \Rightarrow \text{subListAllDifferent}(L)}$$

- **rowsAgg2swi** is defined as follows, where subListAggregate/4 is defined in the CSeE library to aggregate the items among each sublist using given binary operators Bo and Ro.

$$\frac{L1 \leftarrow \text{varListByRows}(CR), \quad L2 \leftarrow \text{varList}(L)}{\text{ssRowsAggregate}(Bo,CR,Ro,L) \Rightarrow \text{subListAggregate}(Bo,L1,Ro,L2)}$$

- **diaAgg2swi** is defined in a similar way for the constraint on diagonal aggregation. So are colsAgg2swi and bdiaAgg2swi for converting the constraints on column aggregation and back diagonals, respectively, to the same subListAggregate/4.

$$\frac{L1 \leftarrow \text{varListByDiagonal}(CR), \quad L2 \leftarrow \text{varList}(L)}{\text{ssDiagonalAggregate}(Bo,CR,Ro,L) \Rightarrow \text{subListAggregate}(Bo,L1,Ro,L2)}$$

- **pairsOp2swi** is for ssPairsOp, where the SWI predicate pairAggregate/5 is defined in the CSeE library to apply the arithmetic operator Bo on each pairwise items in lists L1 and L2, and then connects to the corresponding item in the list L3 by applying the relation operator Ro.

$$\frac{L1 \leftarrow \text{varList}(CR1), \quad L2 \leftarrow \text{varList}(CR2), \quad L3 \leftarrow \text{varList}(CR2)}{\text{ssPairsOp}(CR1,Bo,CR2,Ro,CR3) \Rightarrow \text{pairsAggregate}(L1,Bo,L2,Ro,L3)}$$

**5.3. The CSeE Library for SWI-Prolog**

The CSeE system provides a module file, named "csee.pl", for SWIProlog. The module file contains a set of built-in predicates, such as subListAllDifferent/1 and subListAggregate/4 shown below, which will be loaded automatically when the Prolog engine is started:

```
:- module(csee, [ subListAllDifferent/1,
         subListAggregate/4,
         pairsAggregate/5, ... ]).

subListAllDifferent([ ]).
subListAllDifferent([F | R]) :-
all_different(F),
subListAllDifferent(R).

pairsAggregate([ ], _, [ ], _, _).
pairsAggregate([F1|T1],B,[F2|T2],R,[F|T]) :-
X1 =.. [B,F1,F2],
X2 =.. [R,X1,F],
X2,
```

pairsAggregate(T1,B,T2,R,T).

subListAggregate(B, [ ], _, _).

subListAggregate(B, [F1|T1], R, [F2|T2]) :-
aggregate(B, F1, R, F2),
subListAggregate(B, T1, R, T2).

aggregate(B,[F|T],R,V) :-
aggrConstraint(B,T,F,S),
S1 =.. [R, S, V],
S1.

aggrConstraint(_, [ ], S, S).
aggrConstraint(B,[F|T],S1,S2) :-
S3 =.. [B, S1, F],
aggrConstraint(B, T, S3, S2).

… …
The CSeE system allows users with CLP programming experience to access the CSeE library for pre-defining other useful built-in predicates and adding constraint handling rules [8, 14] for spreadsheet-related constraints. The module/2 directive declares the module name, and the public (i.e., external visible) predicates of the module.

### 5.4. Program Integration

The CSeE system searches automatically two required constraints: ssVarRange(CR1) and ssConstraintRange(CR2), for constructing a complete specification for a constraint satisfaction problem. With the set of all cell variables, specified by the cell range CR1, and the set of constraints, specified by CR2, the CSeE system can easily integrate all the finite-domain cell variables and the defined constraints into an executable CLP program.

The structure of the integrated program is in a form as follows:

:- use_module(library(clpfd)).
:- use_module(library(csee)).
main(L) :- <S1>,
          labeling([<S2>], L).

where <S1> is a set of generated constraints, <S2> is the optimization indicator, and L is a list of cell variables corresponding to the cell range CR1. For each value in both cell ranges CR1 and CR2, the system applies a corresponding constraint transformation rule described in Section 5.2 on the cell value and appends the converted result into the set <S1>. The optimization indicator <S2> is determined by the user constraint ssMin(V) or ssMax(V) if any. Otherwise, <S2> will be empty. The finite-domain solver module and the CSeE module are loaded at the beginning of the program.

International Journal of Programming Languages and Applications ( IJPLA ) Vol.5, No.2, April 2015

## 6. EXAMPLES

We demonstrate how the CSeE system allows the EXCEL user to extend the spreadsheet's ability to solve certain constraint satisfaction problems. The examples include Sudoku, 8-Queen, and a resource allocation problem.

### 6.1. The 8-Queen Problem

The 8-Queen problem deals with finding all the ways of placing eight queens on a square 8 × 8 chess board, in such a way that no two queens are on the same row, column, diagonal, or back-diagonal. We use the cell matrix from A1 to H8 to represent the 8 × 8 chess board and numbers 1 and 0 in the cell to represent a queen and no queen, respectively. Figure 3 shows the Excel interface for the 8-Queen problem.

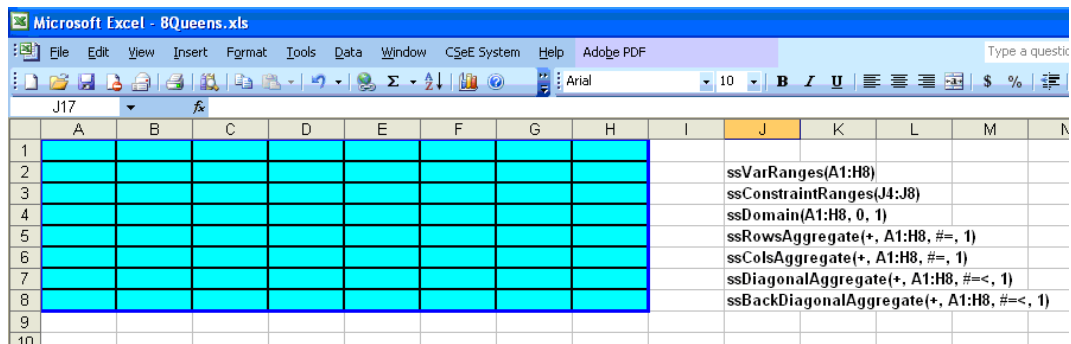

Figure 3: An Excel interface for specifying the 8-Queen problem

The constraints ssVarRange(A1:H8) and ssConstraintRange(J4:J8) are required by the CSeE system to specify the ranges of cell variables and other constraints, respectively. As shown in the figure, we highlighted the cell matrix from A1 to H8 as they are cell variables. The constraint ssDomain(A1:H8, 0, 1) is used to set the domain of the cell variables to a binary domain. ssRowsAggregate(+,A1:H8,#=,1) means that there is one and at most one queen for each row; similar is ssColsAggregate(+,A1:H8,#=,1) for each column. The relational operators #= and #=< are equivalent to = and =<, respectively. The constraint ssDiagonalAggregate(+,A1:H8,#=<,1) means that there is at most one queen for each diagonal; similarly, ssBakDiagonalAggregate(+,A1:H8,#=<,1) is for each back diagonal.

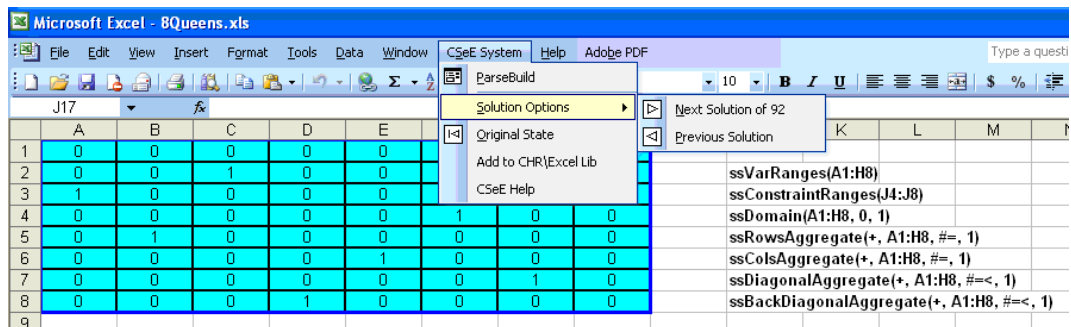

Figure 4: An EXCEL interface after solving the 8-Queen problem

Figure 4 shows an interface where the constraints have been solved. As shown in the figure, the CSeE system provides an add-in menu, named "CSeE system", and a set of sub-menus, which are responsible for communication between the spreadsheet layer and the constraint solving layer:

- The "Parse/Build" menu item performs the following events sequentially: (i) check if the ranges of cell variables and constraints have been specified; (ii) check if the constraints are specified in a correct syntax; (iii) integrate all the inputs into an executable CLP program; (iv) run the program on the SWI-Prolog; and (v) update the spreadsheet properly once answers to the constraint satisfaction problem are generated and returned from the constraint solver.
- The "Solution Options" sub-menu provides the Excel user with the capability to traverse different solutions, where the total number of solutions is indicated as well (e.g., 92 solution for the 8-Queen problem).
- The item "Original State" restores the spreadsheet status before calling the constraint solver so that the user can revise the constraints for new answers.
- The item "Add to CHR\Excel lib" provides experienced CLP users with an interface to update the CSeE library (e.g., adding new built-in predicates or constraint handling rules) for the SWI-Prolog.
- The item "CSeE Help" shows the manual document which introduces the features of the CSeE system and the instructions on how to use the spreadsheet-specific constraint language.

Scalability is an issue we have considered during the design of the SSCL constraint language. For the 8-Queen program, the constraint specification can actually be easily scaled to larger problem (e.g., 20-Queen) with very little change as follows:

ssDomain(A1:T20, 0, 1)
ssRowsAggregate(+, A1:T20, #=, 1)
ssColsAggregate(+, A1:T20, #=, 1)
ssDiagonalAggregate(+, A1:T20, #=<, 1)
ssBackDiagonalAggregate(+, A1:T20, #=<, 1)

### 6.2. The Sudoku Puzzle

The Sudoku puzzle is played in a 9 × 9 grid made up of 3 × 3 sub grids, where each grid contains one instance of digits from 1 to 9. The numbers are entered in such a way that each row, column and region (a 3 × 3 sub grid) has exactly one instance of the digit ranging from 1 to 9. There are digits that are already fixed for some cells, and they act as clues that restrict the solution space in such a way that there is one and only one correct way to populate the remaining regions.

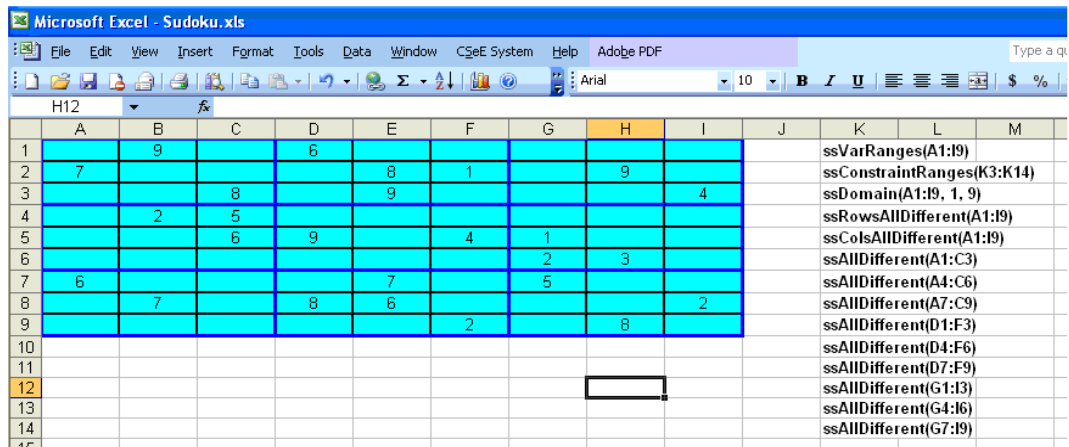

Figure 5: An EXCEL interface for Sudoku

We use the cell matrix from A1 to I9 to represent the 9 × 9 grid board. Figure 5 shows the EXCEL interface for the Sudoku problem. The solution will be immediately shown once the "Parse/Build" is clicked.

### 6.3. Job Assignment

The third example, as shown in Figure 6, is a simple machine/job assignment problem to minimize the total cost. In Figure 6, the machine/job costs are shown in Table 1, and the selection decision will be made in Table 2, including the minimal cost in the cell C16. The minimization is specified by using the constraint ssMin(C16). Once the "Parse/Build" is selected, the CSeE system will return a job assignment solution where the cell variable C16 is minimized. For the example in Figure 6, the minimal cost in the cell C16 will be 22. The order of constraints does not affect the solutions. Users can specify those constraints in any cells except those variable cells.

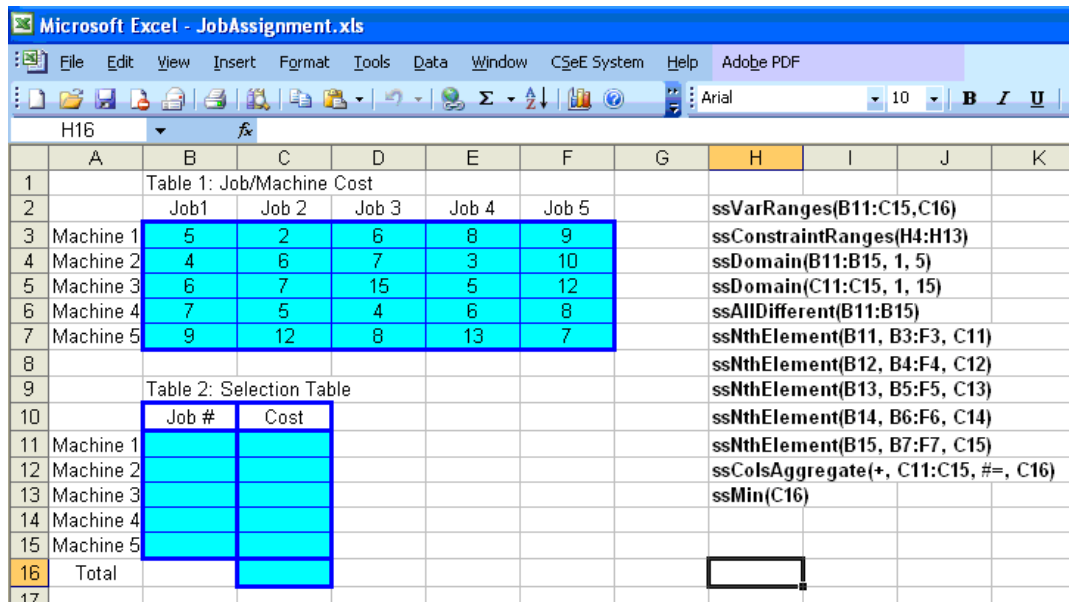

Figure 6: An Excel interface for a job assignment example

## 7. CONCLUSIONS

Spreadsheet users encounter constraint satisfaction problems on a daily basis. As such, the CSeE system is a framework incorporating the spreadsheet paradigm with a constraint solver so that it becomes a useful and indispensable tool for the users. The CSeE's seamless integration with Excel and SWI-Prolog, along with its spreadsheet-specific language, extends the usefulness and usability of the spreadsheet applications, especially for constraint solving problems.

The SSCL spreadsheet-specific constraint language allows users to specify constraints that exist among the cells in a declarative and scalable way in order to solve constraint satisfaction problems. Envisioned as a system that has potential for commercial value, the framework significantly simplifies the complexity involved in developing solutions to many constraint-based application domains from within the Excel's spreadsheet interface. In addition, the CSeE system provides an interface for CLP-experienced users to enhance constraint solving strategies.

Given constraint satisfaction problems encompass a wide array of application domains, the SSCL and the spreadsheet-specific library will be further refined and optimized as applications over these different domains are discovered. Spreadsheet-specific constraint handling rules will be explored to enhance the CSeE system's ability to solve large scale spreadsheet-based applications.

**Authors**

**Ezana N. Beyenne** received his B.S. and M.S. from the Computer Science Department of the University of Nebraska at Omaha in 2006 and 2008, respectively. He has been working in the industry as a software engineer since 2008. Currently he is a Senior Software Engineer working at Farm Credit Services of America.

**Dr. Hai-Feng Guo** received his Ph.D. in Computer Science from New Mexico State University in 2001. He is currently as Associate Professor in the Computer Science Department of the University of Nebraska at Omaha. His research interests focus on declarative programming, constraint logic programming, and software testing.